\documentclass[10pt]{article}
\usepackage{amsmath,amssymb}
\usepackage{url}
\usepackage{graphicx}
\usepackage{subfigure}
\usepackage{cite}
\usepackage{color} 

\newenvironment{myenumerate}{
\begin{enumerate}
 \setlength{\itemsep}{1pt}
 \setlength{\parskip}{0pt}
 \setlength{\parsep}{0pt}}{\end{enumerate}
}

\newenvironment{myitemize}{
\begin{itemize}
 \setlength{\itemsep}{1pt}
 \setlength{\parskip}{0pt}
 \setlength{\parsep}{0pt}}{\end{itemize}
}

\usepackage[labelfont=bf,labelsep=period,justification=raggedright]{caption}

\makeatletter
\renewcommand{\@biblabel}[1]{\quad#1.}
\makeatother

\date{}

\begin{document}

\begin{flushleft}
{\Large
\textbf{Calculating Kolmogorov Complexity from the Output Frequency Distributions of Small Turing Machines}
}
\\
Fernando Soler-Toscano$^{1,5}$, 
Hector Zenil$^{2,5,\ast}$, 
Jean-Paul Delahaye$^{3,5}$
Nicolas Gauvrit$^{4,5}$
\\
\bf{1} Grupo de L\'ogica, Lenguaje e Informaci\'on, Universidad de Sevilla, Sevilla, Spain
\\
\bf{2} Unit of Computational Medicine, Karolinska Institutet, Stockholm, Sweden
\\
\bf{3} Laboratoire d'Informatique Fondamentale de Lille, Universit\'e de Lille I, Lille, France
\\
\bf{4} LDAR, Universit\'e de Paris VII, Paris, France
\\
\bf{5} Algorithmic Nature Group, LABORES, Paris, France
\\
$\ast$ Corresponding author e-mail: hector.zenil@algorithmicnaturelab.org.\\F.S-T. and H.Z. contributed equally.
\end{flushleft}

\date{}



\section*{Abstract}
Drawing on various notions from theoretical computer science, we present a novel numerical approach, motivated by the notion of algorithmic probability, to the problem of approximating the Kolmogorov-Chaitin complexity of short strings. The method is an alternative to the traditional lossless compression algorithms, which it may complement, the two being serviceable for different string lengths. We provide a thorough analysis for all $\sum_{n=1}^{11} 2^n$ binary strings of length $n<12$ and for most strings of length $12\leq n \leq16$ by running all $\sim 2.5 \times 10^{13}$ Turing machines with 5 states and 2 symbols ($8\times 22^9$ with reduction techniques) using the most standard formalism of Turing machines, used in for example the Busy Beaver problem. We address the question of stability and error estimation, the sensitivity of the continued application of the method for wider coverage and better accuracy, and provide statistical evidence suggesting robustness. As with compression algorithms, this work promises to deliver a range of applications, and to provide insight into the question of complexity calculation of finite (and short) strings.\\

Additional material can be found at the \emph{Algorithmic Nature Group} website at\\ \url{http://www.algorithmicnature.org}. An Online Algorithmic Complexity Calculator implementing this technique and making the data available to the research community is accessible at \url{http://www.complexitycalculator.com}.\\

Keywords: Algorithmic randomness; Algorithmic probability; Levin's Universal Distribution; Solomonoff induction; Algorithmic Coding theorem; Invariance theorem; Busy Beaver functions; Small Turing machines.

\section*{Introduction}

The evaluation of the complexity of finite sequences is key in many areas of science. For example, the notions of structure, simplicity and randomness are common currency in biological systems epitomized by a sequence of fundamental nature and utmost importance: the DNA. Nevertheless, researchers have for a long time avoided any practical use of the current accepted mathematical theory of randomness, mainly because it has been considered to be useless in practice~\cite{chaitinestonia}. Despite this belief, related notions such as lossless uncompressibility tests have proven relative success, in areas such as sequence pattern detection~\cite{rivals} and have motivated distance measures and classification methods~\cite{cilibrasi} in several areas (see~\cite{li} for a survey), to mention but two examples among many others of even more practical use. The method presented in this paper aims to provide sound directions to explore the feasibility and stability of the evaluation of the complexity of strings by means different to that of lossless compressibility, particularly useful for short strings. The authors known of only two similar attempts to compute the uncomputable, one related to the estimation of a Chaitin Omega number~\cite{calude3}, and of another seminal related measure of complexity, Bennett's Logical Depth~\cite{zenilld,computability}. This paper provides an approximation to the output frequency distribution of all Turing machines with 5 states and 2 symbols which in turn allow us to apply a central theorem in the theory of algorithmic complexity based in the notion of algorithmic probability (also known as Solomonoff's theory of inductive inference) that relates frequency of production of a string and its Kolmogorov complexity hence providing, upon application of the theorem, numerical estimations of Kolmogorov complexity by a method different to lossless compression algorithms.

A previous result~\cite{delahayezenil} using a simplified version of the  method reported here soon found an application in the study of economic time series~\cite{zenileco,linma}, but wider application was preempted by length and number of strings. Here we significantly extend~\cite{delahayezenil} in various directions: (1) longer, and therefore a greater number---by a factor of three orders of magnitude---of strings are produced and thoroughly analyzed; (2) in light of the previous result, the new calculation allowed us to compare frequency distributions of sets from considerable different sources and of varying sizes (although the smaller is contained in the larger set, it is of negligible size in comparison)---they could have been of different type, but they are not (3) we extend the method to sets of Turing machines whose Busy Beaver has not yet been found by proposing an informed method for estimating a reasonably non-halting cutoff value based on theoretical and experimental considerations, thus (4) provide strong evidence that the estimation and scaling of the method is robust and much less dependent of Turing machine sample size, fully quantified and reported in this paper. The results reported here, the data released with this paper and the online program in the form of a calculator, have now been used in a wider number of applications ranging from psychometrics~\cite{gauvrit} to the theory of cellular automata~\cite{zenilca,chaos}, graph theory and complex networks~\cite{zenilgraphs}. In sum this paper provides a thorough description of the method, a complete statistical analysis of the \emph{Coding theorem method} and an online application for its use and exploitation. The calculation presented herein will remain the best possible estimation for a measure of a similar nature with the technology available to date, as an exponential increase of computing resources will improve the length and number of strings produced only linearly if the same standard formalism of Turing machines used is followed.

\section*{Preliminaries}

\subsection*{Kolmogorov complexity}
\label{kolmo}

Central to AIT is the definition of algorithmic (Kolmogorov-Chaitin or program-size) complexity~\cite{kolmo,chaitin}:

\begin{equation}
\label{kolmoeq}
K_T(s) = \min \{|p|, T(p)=s\}
\end{equation}

\noindent where $p$ is a program that outputs $s$ running on a universal Turing machine $T$. A technical inconvenience of $K$ as a function taking $s$ to the length of the shortest program that produces $s$ is its uncomputability. In other words, there is no program which takes a string $s$ as input and produces the integer $K(s)$ as output. This is usually considered a major problem, but one ought to expect a universal measure of complexity to have such a property. The measure was first conceived to define randomness and is today the accepted objective mathematical measure of complexity, among other reasons because it has been proven to be mathematically robust (by virtue of the fact that several independent definitions converge to it). 
If the shortest program $p$ producing $s$ is larger than $|s|$, the length of $s$, then $s$ is considered random. One can approach $K$ using compression algorithms that detect regularities in order to compress data. The value of the compressibility method is that the compression of a string as an approximation to $K$ is a sufficient test of non-randomness. 

It was once believed that AIT would prove useless for any real world applications~\cite{chaitinestonia}, despite the beauty of its mathematical results (e.g. a derivation of G\"odel's incompleteness theorem~\cite{chaitin2}). This was thought to be due to uncomputability and to the fact that the theory's founding theorem (the invariance theorem), left finite (short) strings unprotected against an additive constant determined by the arbitrary choice of programming language or universal Turing machine (upon which one evaluates the complexity of a string), and hence unstable and extremely sensitive to this choice. 

Traditionally, the way to approach the algorithmic complexity of a string has been by using lossless compression algorithms. The result of a lossless compression algorithm is an upper bound of algorithmic complexity. However, short strings are not only difficult to compress in practice, the theory does not provide a satisfactory answer to all questions concerning them, such as the Kolmogorov complexity of a single bit (which the theory would say has maximal complexity because it cannot be further compressed). To make sense of such things and close this theoretical gap we devised an alternative methodology~\cite{delahayezenil} to compressibility for approximating the complexity of short strings, hence a methodology applicable in many areas where short strings are often investigated (e.g. in bioinformatics). This method has yet to be extended and fully deployed in real applications, and here we take a major step towards full implementation, providing details of the method as well as a thorough theoretical analysis.

\subsection*{Invariance and compression}

A fair compression algorithm is one that transforms a string into two components. The first of these is the compressed version while the other is the set of instructions for decompressing the string. Both together account for the final length of the compressed version. Thus the compressed string comes with its own decompression instructions. Paradoxically, lossless compression algorithms are more stable the longer the string. In fact the invariance theorem guarantees that complexity values will only diverge by a constant $c$ (e.g. the length of a compiler, a translation program between $U_1$ and $U_2$) and will converge at the limit. \\

\noindent \textbf{Invariance Theorem}~\cite{calude,li}: If $U_1$ and $U_2$ 
are two universal Turing machines and $K_{U_1}(s)$ and $K_{U_2}(s)$ the
algorithmic complexity of $s$ for $U_1$ and $U_2$, there exists a constant $c$ such that:  

\begin{equation}
\label{invariance}
| K_{U_1}(s) - K_{U_2}(s) | < c
\end{equation}

Hence the longer the string, the less important the constant $c$ or choice of programming language or universal Turing machine. However, in practice $c$ can be arbitrarily large, thus having a very great impact on finite short strings.  Indeed, the use of data lossless compression algorithms as a method for approximating the Kolmogorov complexity of a string is accurate in direct proportion to the length of the string.

\subsection*{Solomonoff-Levin Algorithmic Probability}
\label{coding}

The algorithmic probability (also known as Levin's semi-measure) of a string $s$ is a measure that describes the expected probability of a random program $p$ running on a universal (prefix-free\footnote{The group of valid programs forms a prefix-free set, that is no element is a prefix of any other, a property necessary to keep $0 < \mathfrak{m}(s) < 1$. For details see~\cite{calude}.}) Turing machine $T$ producing $s$. Formally~\cite{solomonoff,levin,chaitin}, 

\begin{equation}
\label{codingeq}
\mathfrak{m}(s) = \sum_{p:T(p) = s} 1/2^{|p|}
\end{equation}

Levin's semi-measure\footnote{It is called \emph{semi} measure because the sum is not always 1, unlike probability measures. This is due to the Turing machines that never halt.} $\mathfrak{m}(s)$ defines  the so-called Universal Distribution~\cite{kircher}. Here we propose to use $\mathfrak{m}(s)$ as an alternative to the traditional use of compression algorithms to calculate $K(s)$ by means of the following theorem.\\

\noindent \textbf{Coding Theorem}~\cite{levin}: There exists a constant $c$ such that: 
\begin{equation}
|-\log_2 \mathfrak{m}(s) - K(s)| < c
\end{equation}

That is, if a string has many long descriptions it also has a short one~\cite{cover}. It beautifully connects frequency to complexity---the frequency (or probability) of occurrence of a string with its algorithmic (Kolmogorov) complexity. The coding theorem implies that~\cite{delahayezenil} one can calculate the Kolmogorov complexity of a string from its frequency~\cite{zenil2007,delahaye2007,thesis,delahayezenil}, simply rewriting the formula as:

\begin{equation}
K(s)=-\log_2 \mathfrak{m}(s) + O(1)
\end{equation}

An important property of $\mathfrak{m}$ as a semi-measure is that it dominates any other effective semi-measure $\mu$ because there is a constant $c_\mu \neq 0$ such that for all $s$, $\mathfrak{m}(s) \geq c_\mu\mu(s)$ hence called \emph{Universal}~\cite{kircher}.

\subsection*{The Busy Beaver function}

\noindent \textbf{Notation:} We denote by $(n,2)$ the class (or space) of all $n$-state 2-symbol Turing machines (with the halting state not included among the $n$ states).\\

In addressing the problem of approaching $m(s)$ by running computer programs (in this case deterministic Turing machines) one can use the known values of the so-called Busy Beaver functions as suggested by and used in~\cite{thesis,delahayezenil}. The Busy Beaver functions $\sum(n,m)$ and $S(n,m)$ can be defined as follows: \\

\noindent \textbf{Busy Beaver functions} (Rado~\cite{rado}): If $\sigma_T$ is
the number of `1s' on the tape of a Turing machine $T$ with $n$ states
and $m$ symbols upon halting starting from a blank
  tape (no input), then the Busy Beaver function $\sum(n,m)=\max{\{\sigma_T :
  T\in(n,m) \normalsize{\textbf{ }T\textbf{
    }halts}\}}$. Alternatively, if $t_T$ is the number of steps that a
machine $T$ takes before halting from a blank tape, then
$S(n,m)=\max{\{t_T : T\in(n,m) \normalsize{\textbf{ }T\textbf{
    }halts}\}}$.\\ 

In other words, the Busy Beaver functions are the functions that return the longest written tape and longest runtime in a set of Turing machines with $n$ states and $m$ symbols. $\sum(n, m)$ and $S(n, m)$ are noncomputable functions by reduction to the halting problem. In fact $\sum(n, m)$ grows faster than any computable function can. Nevertheless, exact values can be calculated for small $n$ and $m$, and they are known for, among others, $m=2$ symbols and $n < 5$ states. A program showing the evolution of all known Busy Beaver machines developed by one of this paper's authors is available online~\cite{busyhz}. 

\begin{figure}[h!]
\centering
\scalebox{.36}{\includegraphics{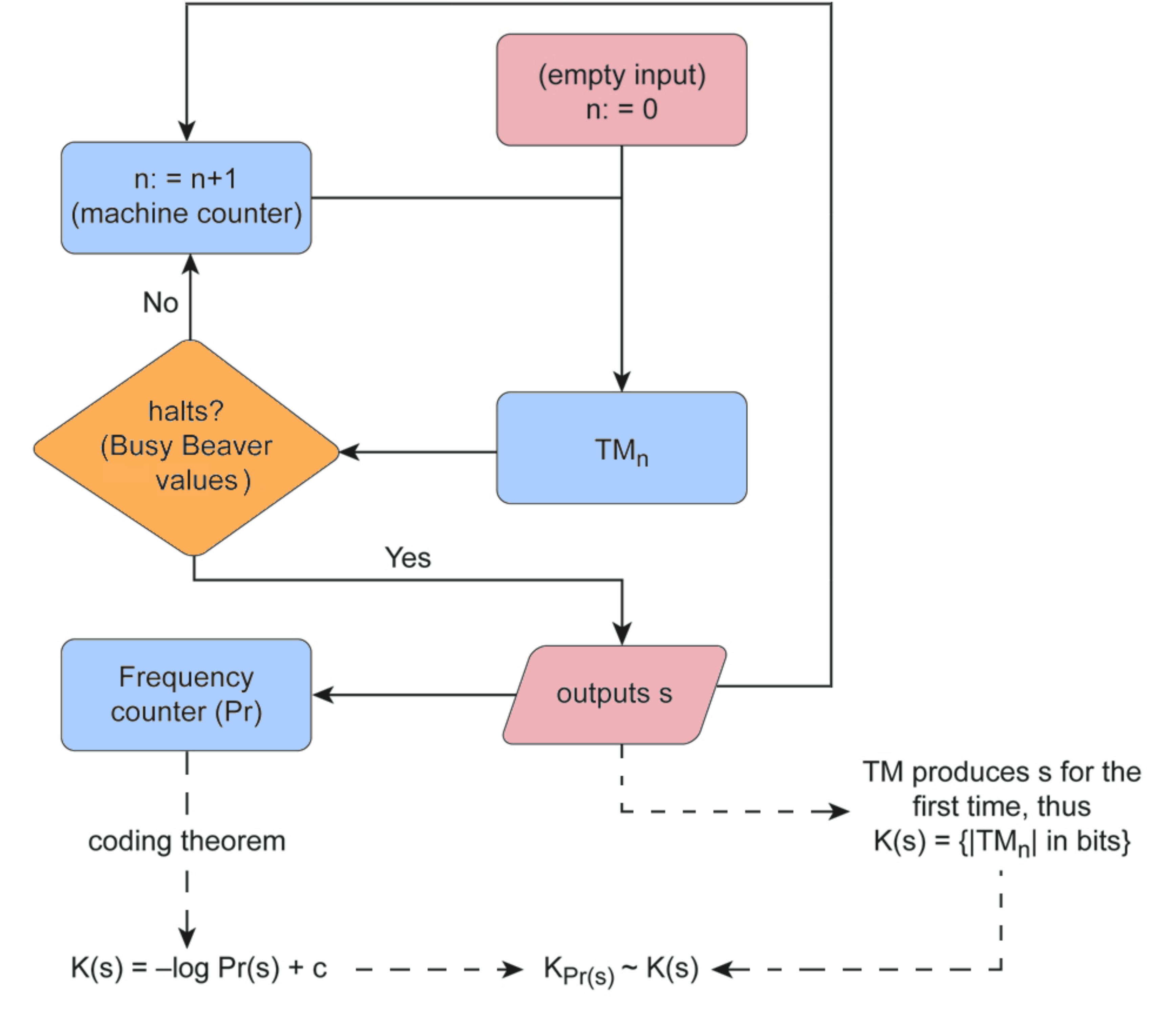}}
\caption{\label{flowchart}A flow chart illustrating the \emph{Coding Theorem Method}, a never-ending algorithm for evaluating the (Kolmogorov) complexity of a (short) string making use of several concepts and results from theoretical computer science, in particular the halting probability, the Busy Beaver problem, Levin's semi-measure and the Coding theorem. The Busy Beaver values can be used up to 4 states for which they are known, for more than 4 states an informed maximum runtime is used as described in this paper, informed by theoretical~\cite{calude2} and experimental (Busy Beaver values) results. Notice that $Pr$ are the probability values calculated dynamically by running an increasing number of Turing machines. $Pr$ is intended to be an approximation to $\mathfrak{m}(s)$ out of which we build $D(n)$ after application of the Coding theorem.}
\end{figure}

This allows one to circumvent the problem of noncomputability for small Turing machines of the size that produce short strings whose complexity is approximated by applying the algorithmic Coding theorem (see Fig.~\ref{flowchart}). As is widely known, the Halting problem for Turing machines is the problem of deciding whether an arbitrary Turing machine $T$ eventually halts on an arbitrary input $s$. Halting computations can be recognized by running them for the time they take to halt. The problem is to detect non-halting programs, programs about which one cannot know in advance whether they will run forever or eventually halt.

\subsection*{The Turing machine formalism}
\label{formal}

It is important to describe the Turing machine formalism because numerical values of algorithmic probability for short strings will be provided under this chosen standard model of a Turing machine.\\

Consider a Turing machine with the binary alphabet $\sum=\{0,1\}$ and $n$ states $\{1,2, \ldots n\}$ and an additional Halt state denoted by 0 (as defined by Rado in his original Busy Beaver paper~\cite{rado}).\\

The machine runs on a $2$-way unbounded tape. At each step: 

\begin{myenumerate}
\item the machine's current ``state''; and
\item the tape symbol the machine's head is scanning
\end{myenumerate}

define each of the following: 

\begin{myenumerate}
\item a unique symbol to write (the machine can overwrite a $1$ on a $0$, a $0$ on a $1$, a $1$ on a $1$, and a $0$ on a $0$);
\item a direction to move in: $-1$ (left), $1$ (right) or $0$ (none, when halting); and
\item a state to transition into (which may be the same as the one it was in).
\end{myenumerate}

The machine halts if and when it reaches the special halt state 0. There are $(4n + 2)^{2n}$ Turing machines with $n$ states and 2 symbols according to the formalism described above. The output string is taken from the number of contiguous cells on the tape the head of the halting $n$-state machine has gone through. A machine produces a string upon halting.

\section*{Methodology}
\label{D}

One can attempt to approximate $\mathfrak{m}(s)$ (see Eq.~3) by running every Turing machine an particular enumeration, for example, a quasi-lexicographical ordering, from shorter to longer (with number of states $n$ and 2 fixed symbols). It is clear that in this fashion once a machine produces $s$ for the first time, one can directly calculate an approximation of $K$, because this is the length of the first Turing machine in the enumeration of programs of increasing size that produces $s$. But more important, one can apply the Coding theorem to extract $K(s)$ from $\mathfrak{m}(s)$ directly from the output distribution of halting Turing machines. Let's formalize this by using the function $D(n, m)$ as the function that assigns to every string $s$ produced in $(n, m)$ the quotient: (number of times that a machine in $(n, m)$ produces $s$) / (number of machines that halt in $(n, m)$) as defined in~\cite{thesis,delahayezenil}. More formally,\\

\begin{equation}
\label{d}
D(n, m)(s)=\frac{|\{T\in(n, m) : T(p) = s\}|}{|\{T\in(n, m) : T \textit{ halts }\}|}
\end{equation}

Where $T(p)$ is the Turing machine with number $p$ (and empty input) that produces $s$ upon halting and $|A|$ is, in this case, the cardinality of the set $A$. A variation of this formula closer to the definition of $\mathfrak{m}$ is given by:

\begin{equation}
\label{d2}
D^\prime(n, m)(s)=\frac{|\{T\in(n, m) : T(p) = s\}|}{|\{T\in(n, m)\}|}
\end{equation}

$D^\prime$ is strictly smaller than 1 for $m,n\rightarrow \infty$, because of the Turing machines that never halt, just as it occurs for $\mathfrak{m}$. However, for fixed $n$ and $m$ the sum of $D$ will always be 1. We will use Eq.~\ref{d} for practical reasons, because it makes the frequency values more readable (most machines don't halt, so those halting would have a tiny fraction with too many leading zeros after the decimal). Moreover, the function $(n, m) \rightarrow D(n, m)$ is non-computable~\cite{thesis,delahayezenil} but it can be approximated from below, for example, by running small Turing machines for which known values of the Busy Beaver problem~\cite{rado} are known. For example ~\cite{brady}, for $n=4$, the Busy Beaver function for maximum runtime $S$, tells us that $S(4, 2)=107$, so we know that a machine running on a blank tape will never halt if it hasn't halted after 107 steps, and so we can stop it manually. In what follows we describe the exact methodology. From now on, $D(n)$ with a single parameter will mean $D(n, 2)$.

We call this method the \emph{Coding Theorem Method} to approximate $K$ (which we will denote by $K_\mathfrak{m}$).

\subsection*{Kolmogorov complexity from the output frequency of small Turing machines}

Approximations from the output distribution of Turing machines with 2 symbols and $n=1, \ldots, 4$ states for which the Busy Beaver values are known were estimated before~\cite{thesis,delahayezenil} but for the same reason the method was not scalable beyond $n=4$. The formula for the number of machines given a number of states $n$ is given by $(4n+2)^{2n}$ derived from the formalism described. There are 26\,559\,922\,791\,424 Turing machines with 5 states\footnote{That is, for the reader amusement, about the same number of red cells in the blood of an average adult.}. Here we describe how an optimal runtime based on theoretical and experimental grounds can be calculated to scale the method to larger sets of small Turing machines.

Because there are a large enough number of machines to run even for a small number of machine states ($n$), applying the Coding theorem provides a finer and increasingly stable evaluation of $K(s)$ based on the frequency of production of a large number of Turing machines, but the number of Turing machines grows exponentially, and producing $D(5)$ requires considerable computational resources.

\subsection*{Setting an informed runtime}
\label{sec:setting-runtime}

The Busy Beaver for Turing machines with 4 states is known to be 107 steps~\cite{brady}, that is, any Turing machine with 2 symbols and 4 states running longer than 107 steps will never halt. However, the exact number is not known for Turing machines with 2 symbols and 5 states, although it is believed to be 47\,176\,870, as there is a candidate machine that runs for this long and halts and no machine greater runtime has yet been found.

So we decided to let the machines with 5 states run for 4.6 times the Busy Beaver value for 4-state Turing machines (for 107 steps), knowing that this would constitute a sample significant enough to capture the behavior of Turing machines with 5 states. The chosen runtime was rounded to 500 steps, which was used to build the output frequency distribution for $D(5)$. The theoretical justification for the pertinence and significance of the chosen runtime is provided in the following sections.

\subsection*{Reduction techniques}
\label{sec:some-reductions}

We didn't run all the Turing machines with 5 states to produce $D(5)$ because one can take advantage of symmetries and anticipate some of the behavior of the Turing machines directly from their transition tables without actually running them (this is impossible in general due to the halting problem). We avoided some trivial machines whose results we know without having to run them (reduced enumeration). Also, some non-halting machines were detected before consuming all the runtime (filters). The following are the reductions utilized in order to reduce the number of total machines and therefore the computing time for the approximation of $D(5)$.

\subsection*{Exploiting symmetries}
\label{sec:expl-simm}

\subsubsection*{Symmetry of 0 and 1}
\label{sec:symmetry-symbols}

The blank symbol is one of the 2 symbols (0 or 1) in the first run, while the
 other is used in the second run (in order to avoid any asymmetries due 
to the choice of a single blank symbol). In other words, we considered two runs for each Turing machine, one with 0 as the blank symbol (the symbol with which 
the tape starts out and fills up), and an additional run with 1
 as the blank symbol. This means that every machine was run twice. Due to the
 symmetry of the computation, there is no real need to run each machine
 twice; one can \emph{complete} the string frequencies by assuming that 
each string produced produced by a Turing machine has its complement produced by another symmetric machine with the same frequency, we then group and divide by symmetric groups. We 
used this technique from $D(1)$ to $D(4)$. A more detailed 
explanation of how this is done is provided in~\cite{thesis,delahayezenil} using Polya's counting theorem.

\subsubsection*{Symmetry right-left}
\label{sec:symmetry-right-left}

We can exploit the right-left symmetry. We may,
 for example, run only those machines with an initial transition 
(initial state and blank symbol) moving to the right 
and to a state different from the initial one (because an initial
 transition to the initial state produces a non-halting machine) and
 the halting one (these machines stop in just one step and produce
`0' or `1'). 

For every string produced, we also count the reverse in the 
tables. We count the corresponding number of one-symbol 
strings and non-halting machines as well.

\subsubsection*{Reduction techniques by exploiting symmetries} 
\label{sec:prop-enum-left}

If we consider only machines with a starting transition that moves to 
the right and goes to a state other than the starting and halting 
states, the number of machines is given by
\[total(n) = 2(n-1)\big((4n+2)^{2n-1}\big)\]
Note that for the starting transition there are $2(n-1)$ possibilities
($2$ possible symbols to write and $n-1$ possible new states, as we
exclude the starting and halting states). For the other $2n-1$
transitions there are $4n+2$ possibilities. 

We can make an enumeration from $0$ to $total(n)-1$. Of course, this 
enumeration is not the same as the one we use to explore the 
whole space. The same number will not correspond to the same machine. 

In the whole $D(5)$ space there are $(4n+2)^{2n}$ machines, so it is a 
considerable reduction. This reduction in $D(5)$ means that in the 
reduced enumeration we have $4/11$ of the machines we had in the original 
enumeration.
 

\subsection*{Strings to complete after running the reduced enumeration}
\label{sec:string-compl-after}

Suppose that using the previous enumeration we run $M$ machines for
$D(n)$ with blank symbol $0$. $M$ can be the total
number of machines in the reduced space  
or a random number of machines in it (such as we use to study the runtime distribution, as it is better described below). 

For the starting transition we considered only $2(n-1)$ possibilities out of $4n+2$ possible transitions in the whole space. Then, we proceeded as follows to complete the strings produced by the $M$ runs.
\begin{myenumerate}
\item We avoided $2(n-1)$ transitions moving left to a different
 state than the halting and starting ones. We completed such
 transitions by reversing all the strings found. Non-halting machines
 were multiplied by $2$. 
\item We also avoided $2$ transitions (writing `0' or `1') from
  the initial to the halting state. We completed such transitions by
  \begin{myitemize}
  \item Including $\frac{M}{2(n-1)}$ times `0'.
  \item Including $\frac{M}{2(n-1)}$ times `1'.
  \end{myitemize}
\item Finally, we avoided $4$ transitions from the initial state to
 itself ($2$ movements $\times$ $2$ symbols). We completed by
 including $\frac{2M}{n-1}$ non-halting machines. 
\end{myenumerate}

With these completions, we obtained the output strings for the blank symbol
 $0$. To complete for the blank symbol $1$ we took the complement to $1$ 
of each string produced and counted the non-halting machines twice. 

Then, by running $M$ machines, we obtained a result representing
$\frac{M(4n+2)}{(n-1)}$, that for $n=5$ is $5.5M$. 

\subsection*{Detecting non-halting machines}
\label{sec:detect-non-halt}

It is useful to avoid running machines that we can 
easily check that will not stop. These machines will consume the runtime 
without yielding an output. 

The reduction in the enumeration that we have shown reduces the number
 of machines to be generated. Now we present some reductions that work 
after the machines are generated, in order to detect non-halting 
computations and skip running them. Some of these were detected when filling 
the transition table, others at runtime. 

\subsubsection*{Machines without transitions to the halting state}
\label{sec:mach-with-trans}

While we are filling the transition table, if a certain transition goes to the halting state, we can activate a flag. If after completing the transition table the flag is not activated, we know that the machine won't stop. 

In our reduced enumeration there are $2(n-1)\left((4n)^{2n-1}\right)$ machines of this kind. In $D(5)$ this is $4.096 \times 10^{12}$ machines. It represents 42.41\% of the total number of machines. 

The number of machines in the reduced enumeration that are not filtered as non-halting when filling the transition table is 5\,562\,153\,742\,336. That is 504.73 times the total number of machines that fully produce $D(4)$.

\subsubsection*{Detecting escapees}
\label{sec:detecting-loops}

There should be a great number of escapees, that is, machines that run infinitely in the same direction over the tape. 

Some kinds are simple to check in the simulator. We can use a counter that indicates the number of consecutive not-previously-visited tape positions that the machines visits. If the counter exceeds the number of states, then we have found a loop that will repeat infinitely. To justify this, let us ask you to suppose that at some stage the machine is visiting a certain tape-position for the first time, moving in a specific direction (the direction that points toward new cells). If the machine continues moving in the same direction for $n+1$ steps, and thus reading blank symbols, then it has repeated some state $s$ in two transitions. As it is always reading (not previously visited) blank symbols, the machine has repeated the transition for $(s,b)$ twice, $b$ being the blank symbol. But the behavior is deterministic, so if the machine has used the transition for $(s,b)$ and after some steps in the same direction visiting blank cells, it has repeated the same transition, it will continue doing so forever, because it will always find the same symbols. 

There is another possible direction in which this filter may apply: if the symbol read is a blank one not previously visited, the shift is in the direction of new cells and there is no modification of state. In fact this would be deemed an escapee, because the machine runs for $n+1$ new positions over the tape. But it is an escapee that is especially simple to detect, in just one step and not $n+1$. We call the machines detected by this simple filter ``short escapees'', to distinguish them from other, more general escapees. 

\subsubsection*{Detecting cycles}
\label{sec:detecting-cycles}

We can detect cycles of period two. They are 
produced when in steps $s$ and $s+2$ the tape is identical and the 
machine is in the same state and the same position. When this is
 the case, the cycle will be repeated infinitely. To detect it, 
we have to anticipate the following transition that will
apply in some cases. In a cycle of period two, the machine cannot
change any symbol 
 on the tape, because if it did, the tape would be 
different after two steps. Then the filter would be activated 
when there is a transition that does not change the tape, for instance 
\[\{s,k\}\to\{s',k,d\}\]
where $d\in\{-1,1\}$ is some direction (left, right) and the head is 
at position $i$ on tape $t$, which is to say, reading the symbol $t[i]$. Then, 
there is a cycle of period two if and only if the transition that 
corresponds to $\{s',t[i+d]\}$ is 
\[\{s',t[i+d]\}\to\{s,t[i+d],-d\}\]


\subsection*{Number of Turing machines}
\label{sec:checking-filters}

We calculated $D(4)$ with and without all the filters as suggested in~\cite{delahayezenil}. Running $D(4)$ without reducing the number or detecting non-halting machines took 952 minutes. Running the reduced enumeration with non-halting detectors took 226 minutes. 

We filtered the following non-halting machines:
\begin{center}
   \tabcolsep=0.13cm
  \begin{tabular}{|l|r|}
  \hline
  \textbf{Filter} & \textbf{number of TMs} \\\hline\hline
  \small  machines without transitions to the halting state & 1\,610\,612\,736
  \\
 \small short escapees & 464\,009\,712 \\
\small other escapees & 336\,027\,900 \\
\small cycles of period two  & 15\,413\,112 \\
\small machines that consume all the runtime & 366\,784\,524\\
  \textbf{Total} & 2792847984\\\hline
\end{tabular}
\end{center}

Running all the Turing machines with 5 states in the reduced 
enumeration up to 500 steps for the calculation of $D(5)$ took 18 days using 25 x86-64 CPUs running at 2128 MHz with 4 GB of memory each\footnote{A supercomputer located at the Centro Inform\'atico Cient\'ifico de Andaluc\'ia (CICA), Spain.}. In order to save space in the output of $D(5)$, our C++ simulator produced partial results every $10^9$ consecutive machines according to the enumeration. Every  $10^9$ machines, the counters for each string produced were 
updated. The final classification is only 4.1 Megabytes but we can estimate
 the size of the output had we not produced partial results on the 
order of 1.28 Terabytes for the reduced space and 6.23 Terabytes for
 the full one. If we were to include in the output an indication for 
non-halting machines, the files would grow an extra 1.69 Terabytes for
 the reduced enumeration and 8.94 Terabytes for the full one.

\section*{Results}

Samples of strings extracted from the output frequency of these machines are shown in Tables~\ref{main}, \ref{d57}, \ref{climbers}, \ref{largest}, \ref{trans}, \ref{trans2} and  \ref{transb2}; highlighting various important features found in the distributions. Table~\ref{main} provides a glance at $D(5)$ showing the 147 most frequent (and therefore simplest) calculated strings out of
99\,608. The top strings of $D(5)$ conform to an intuition of simplicity. Table~\ref{d57} shows all the $2^n$ strings for $n=7$, hence displaying what $D(5)$ suggests are the strings sorted from lowest to highest complexity, which seems to agree well with the intuition of simple (from top left) to random-looking (bottom right).

\begin{center}
\begin{table}[h]
\small
   \caption{\label{main} The 147 most frequent strings from $D(5)$ (by row). The first column is a counter to help locate the rank of each string.}
   \begin{center}
      \tabcolsep=0.12cm
   \begin{tabular}{|r|c|c|c|c|c|c|c|}
   \hline
 1 & 1 & 0 & 11 & 10 & 01 & 00 & 111 \\
 \hline
 8 & 000 & 110 & 100 & 011 & 001 & 101 & 010 \\
 \hline
  15 & 1111 & 0000 & 1110 & 1000 & 0111 & 0001 & 1101 \\
  \hline
  22 & 1011 & 0100 & 0010 & 1010 & 0101 & 1100 & 0011 \\
  \hline
 29 & 1001 & 0110 & 11111 & 00000 & 11110 & 10000 & 01111 \\
  \hline
  36 & 00001 & 11101 & 10111 & 01000 & 00010 & 11011 & 00100 \\
  \hline
  43 & 10110 & 10010 & 01101 & 01001 & 10101 & 01010 & 11010 \\
  \hline
  50 & 10100 & 01011 & 00101 & 11100 & 11000 & 00111 & 00011 \\
  \hline
  57 & 11001 & 10011 & 01100 & 00110 & 10001 & 01110 & 111111 \\
  \hline
  64 & 000000 & 111110 & 100000 & 011111 & 000001 & 111101 & 101111 \\
  \hline
  71 & 010000 & 000010 & 101010 & 010101 & 101101 & 010010 & 111011 \\
  \hline
  78 & 110111 & 001000 & 000100 & 110101 & 101011 & 010100 & 001010 \\
  \hline
  85 & 101001 & 100101 & 011010 & 010110 & 110110 & 100100 & 011011 \\
  \hline
  92 & 001001 & 111100 & 110000 & 001111 & 000011 & 101110 & 100010 \\
  \hline
  99 & 011101 & 010001 & 110010 & 101100 & 010011 & 001101 & 111001 \\
  \hline
  106 & 100111 & 011000 & 000110 & 111010 & 101000 & 010111 & 000101 \\
  \hline
  113 & 100110 & 011001 & 110011 & 001100 & 100001 & 011110 & 110100 \\
  \hline
  120 & 001011 & 111000 & 000111 & 110001 & 100011 & 011100 & 001110 \\
  \hline
  127 & 1111111 & 0000000 & 1111110 & 1000000 & 0111111 & 0000001 & 1010101 \\
  \hline
  134 & 0101010 & 1111101 & 1011111 & 0100000 & 0000010 & 1111011 & 1101111 \\
  \hline
  141 & 0010000 & 0000100 & 1110111 & 0001000 & 1111100 & 1100000 & 0011111 \\
\hline
   \end{tabular}
   \end{center}
\end{table}
\end{center}

\begin{center}
\begin{table}[h!]
\small
   \caption{\label{d57} All the $2^n$ strings for $n=7$ from $D(5)$ sorted from highest frequency (hence lowest complexity) to lowest frequency (hence highest (random) complexity). Strings in each row have the same frequency (hence the same Kolmogorov complexity). There are 31 different groups representing the different complexities of the $2^7=128$ strings.}
   \begin{center}
   \tabcolsep=0.12cm
   \begin{tabular}{|r|c|c|c|c|c|c|c|c|}
   \hline
 1&1111111 & 0000000 &  &  &  &  &  &  \\
 \hline
 2&1111110 & 1000000 & 0111111 & 0000001 &  &  &  &  \\
 \hline
 3&1010101 & 0101010 &  &  &  &  &  &  \\
 \hline
 4&1111101 & 1011111 & 0100000 & 0000010 &  &  &  &  \\
 \hline
 5&1111011 & 1101111 & 0010000 & 0000100 &  &  &  &  \\
 \hline
 6&1110111 & 0001000 &  &  &  &  &  &  \\
 \hline
 7&1111100 & 1100000 & 0011111 & 0000011 &  &  &  &  \\
 \hline
 8&1011010 & 1010010 & 0101101 & 0100101 &  &  &  &  \\
 \hline
 9&1101101 & 1011011 & 0100100 & 0010010 & 1111001 & 1001111 & 0110000 & 0000110 \\
 \hline
 10&1110101 & 1010111 & 0101000 & 0001010 &  &  &  &  \\
 \hline
 11&1101110 & 1000100 & 0111011 & 0010001 &  &  &  &  \\
 \hline
 12&1101010 & 1010100 & 0101011 & 0010101 &  &  &  &  \\
 \hline
 13&1010110 & 1001010 & 0110101 & 0101001 &  &  &  &  \\
 \hline
 14&1111010 & 1010000 & 0101111 & 0000101 &  &  &  &  \\
 \hline
 15&1110110 & 1001000 & 0110111 & 0001001 &  &  &  &  \\
 \hline
 16&1010001 & 1000101 & 0111010 & 0101110 &  &  &  &  \\
 \hline
 17&1011110 & 1000010 & 0111101 & 0100001 &  &  &  &  \\
 \hline
 18&1011101 & 0100010 &  &  &  &  &  &  \\
 \hline
 19&1101011 & 0010100 & 1001001 & 0110110 &  &  &  &  \\
 \hline
 20&1110011 & 1100111 & 0011000 & 0001100 &  &  &  &  \\
 \hline
 21&1100101 & 1010011 & 0101100 & 0011010 &  &  &  &  \\
 \hline
 22&1011001 & 1001101 & 0110010 & 0100110 & 1000001 & 0111110 &  &  \\
 \hline
 23&1111000 & 1110000 & 0001111 & 0000111 & 1101001 & 1001011 & 0110100 & 0010110 \\
 \hline
 24&1110010 & 1011000 & 0100111 & 0001101 & 1101100 & 1100100 & 0011011 & 0010011 \\
 \hline
 25&1100010 & 1011100 & 0100011 & 0011101 &  &  &  &  \\
 \hline
 26&1100110 & 1001100 & 0110011 & 0011001 &  &  &  &  \\
 \hline
 27&1001110 & 1000110 & 0111001 & 0110001 &  &  &  &  \\
 \hline
 28&1100001 & 1000011 & 0111100 & 0011110 &  &  &  &  \\
 \hline
 29&1110001 & 1000111 & 0111000 & 0001110 &  &  &  &  \\
 \hline
 30&1100011 & 0011100 &  &  &  &  &  &  \\
 \hline
 31&1110100 & 1101000 & 0010111 & 0001011 &  &  &  &  \\
\hline
   \end{tabular}
   \end{center}
\end{table}
\end{center}

\subsection*{Reliability of the approximation of $D(5)$}
\label{time}

\begin{figure}[h!]
\begin{center}
{\scalebox{0.34}{\includegraphics{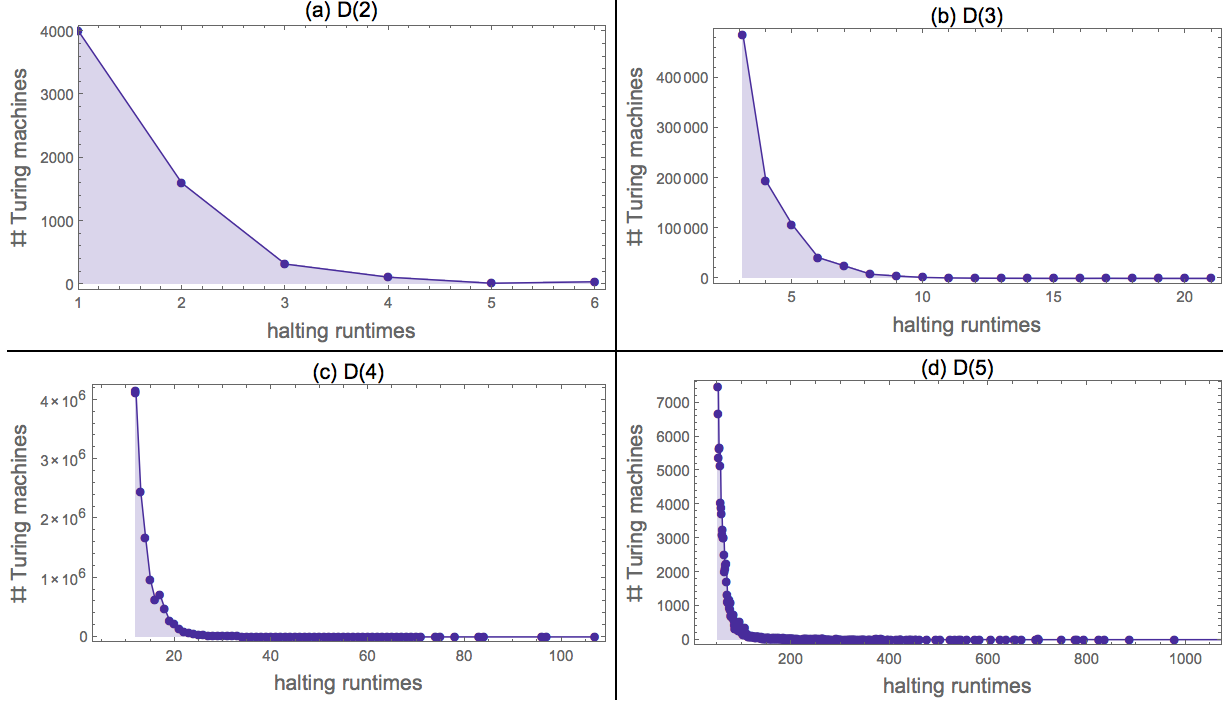}}}
\end{center}
\caption{\label{runtimeplots} Distribution of runtimes from $D(2)$ to $D(5)$. On the $y$-axes are the number of Turing machines and on the $x$-axes the number of steps upon halting. For 5-state Turing machines no Busy Beaver values are known, hence $D(5)$ (Fig. d) was produced by Turing machines with 5 states that ran for at most $t=5000$ steps. These plots show, however, that the runtime cutoff $t=500$ for the production of $D(5)$ covers most of the halting Turing machines when taking a sample of $10^5$ machines letting them run for up to $t=5000$ steps, hence the missed machines in $D(5)$ must be a negligible number for $t=500$.}
\end{figure}

Not all 5-state Turing machines have been used to build $D(5)$, since only the output of machines that halted at or before 500 steps were taken into consideration.
 As an experiment to see how many machines we were leaving out, we ran 
$1.23 \times 10^{10}$ Turing machines for up to 5000 steps (see Fig.~\ref{runtimeplots}d). Among
 these, only 50 machines halted after 500 steps and before 5000 (that 
is less than $1.75164\times 10^{-8}$ because in the reduced 
enumeration we don't include those machines that halt in one step or 
that we know won't halt before generating them, so it is a 
smaller fraction), with the remaining 1\,496\,491\,379 machines not halting 
at 5000 steps. As far as these are concerned--and given the unknown values for the Busy
 Beavers for 5 states--we do not know after how many steps they would
 eventually halt, if they ever do. According to the following analysis, our
 election of a runtime of 500 steps therefore provides a good estimation of
$D(5)$.  

The frequency of runtimes of (halting) Turing machines has theoretically been proven to drop exponentially~\cite{calude2}, and our experiments are closer to the theoretical behavior (see Fig.~\ref{runtimeplots}). To estimate the fraction of halting machines that were missed because Turing machines with 5 states were stopped after 500 steps, we hypothesize that the number of steps $S$ a random halting machine needs before halting is an exponential RV (random variable), defined by $\forall k\geq 1, P(S=k)\propto e^{-\lambda k}.$ We do not have direct access to an evaluation of $P(S=k)$, since we only have data for those machines for which $S\leq 5000$. But we may compute an approximation of $P(S=k | S\leq 5000)$, $1\leq k \leq 5000$, which is proportional to the desired distribution.

\begin{center}
\begin{figure}[h!]
\centering
\includegraphics[width=11.5cm]{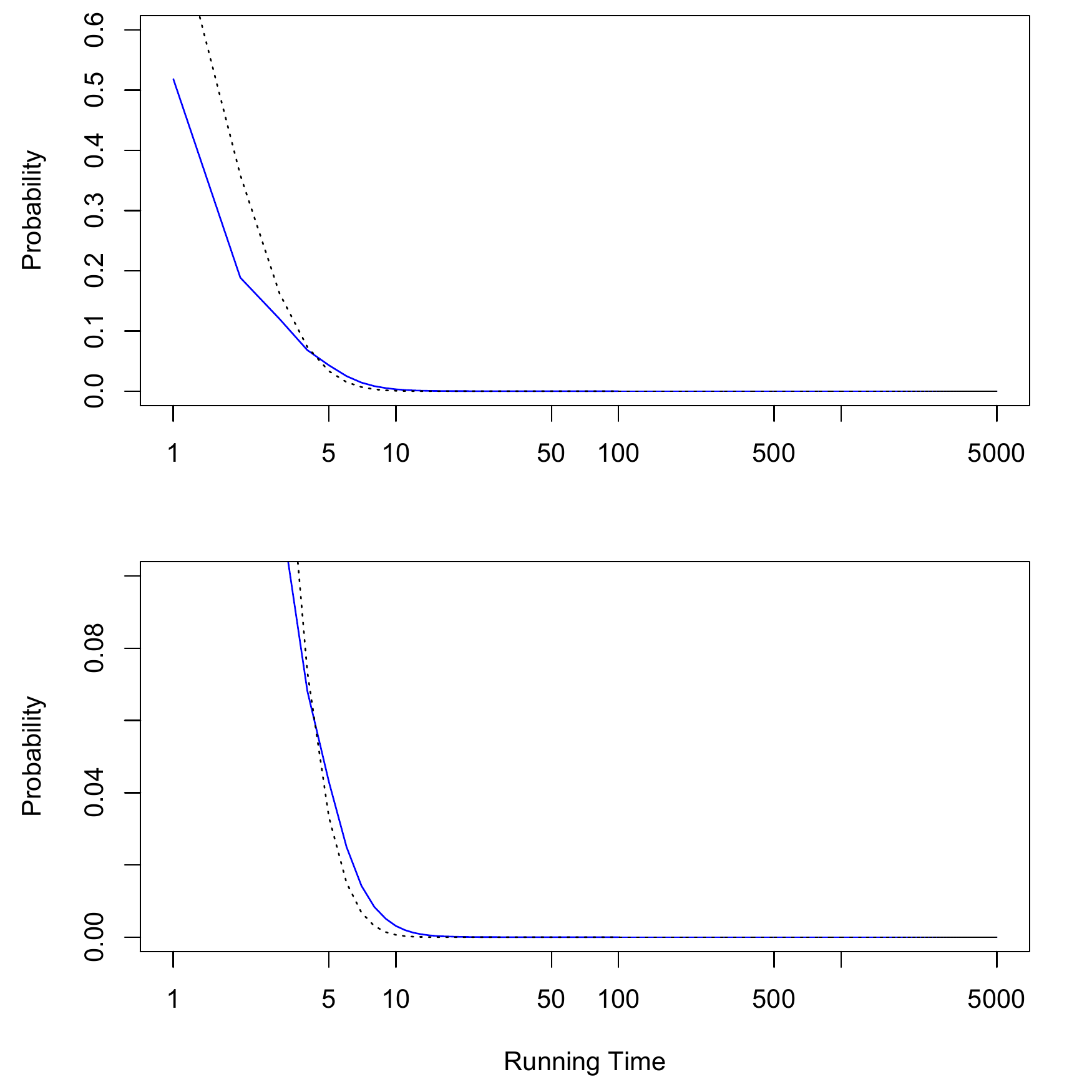}
\caption{\label{runtimeplot} Observed (solid) and theoretical (dotted) $P(S=k|S\leq 5000)$ against $k$. The $x$-axis is logarithmic. Two different scales are used on the $y$-axis to allow for a more precise visualization.}
\end{figure}
\end{center}

A non-linear regression using ordinary least-squares gives the approximation $P(S=k | S\leq 5000)=\alpha e^{-\lambda k}$ with $\alpha = 1.12$ and $\lambda = 0.793$. The residual sum-of-squares is $3.392\times 10^{-3}$, the number of iterations 9 with starting values $\alpha =0.4$ and $\lambda =0.25$.
Fig.~\ref{runtimeplot} helps to visualize how the model fits the data.

 The model's $\lambda$ is the same $\lambda$ appearing in the general law $P(S=k)$, and may be used to estimate the number of machines we lose by using a 500 step cut-off point for running time: $P(k> 500)\approx e^{-500\lambda}\approx 6\times 10^{-173}$. This estimate is far below the point where it could seriously impair our results: the less probable (non-impossible) string according to $D(5)$ has an observed probability of $1.13\times 10^{-9}$.

 Although this is only an estimate, it suggests that missed machines are few enough to be considered negligible.

\section*{Features of $D(5)$}

\subsection*{Lengths}

\begin{center}
\begin{figure}[h!]
\centering
\includegraphics[width=9.7cm]{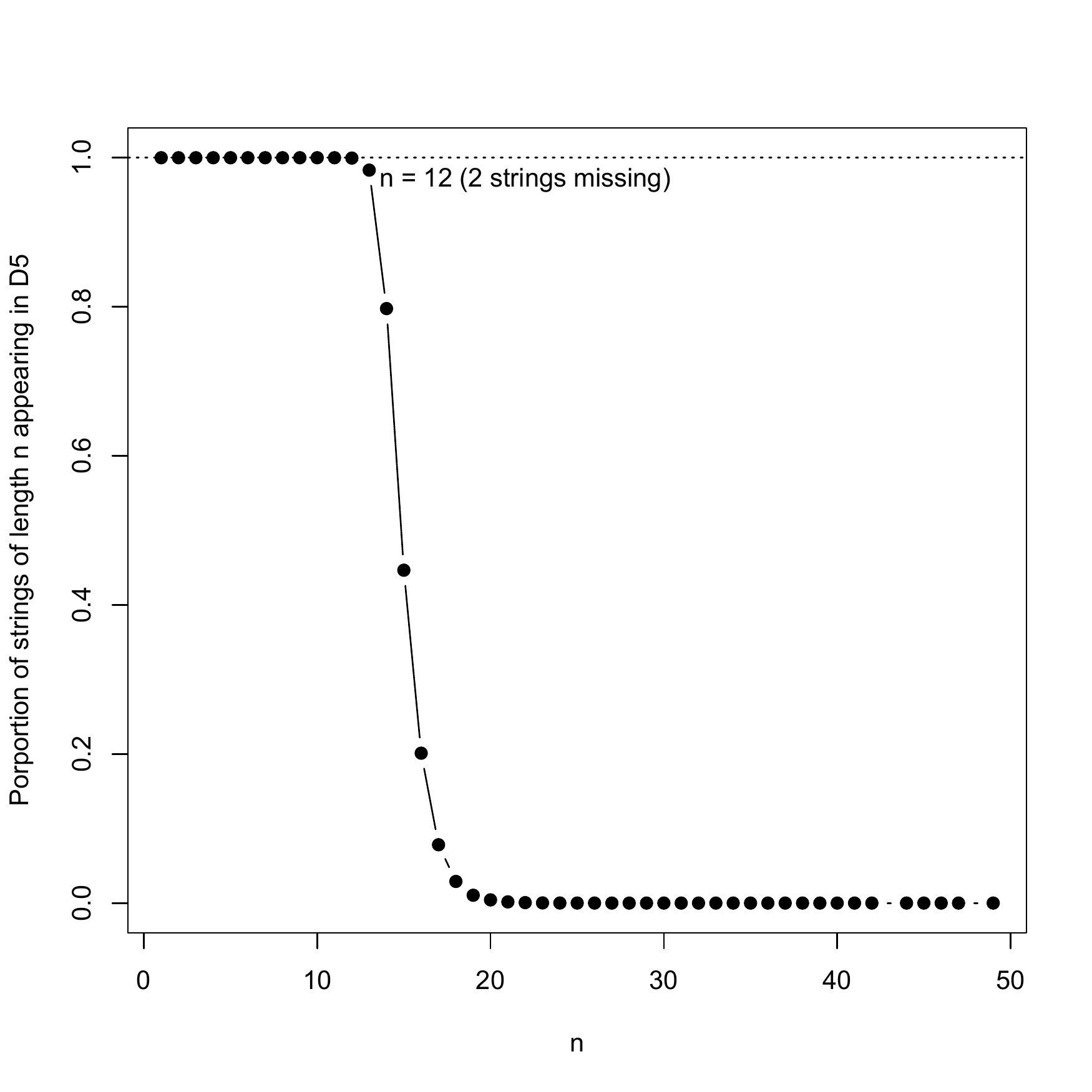}
\caption{\label{PropLengthD(5)} Proportion of all $n$-long strings appearing in $D(5)$ against $n$}
\end{figure}
\end{center}

5-state Turing machines produced 99\,608 different binary strings (to be compared to the 1832 strings for $D(4)$). While the largest string produced for $D(4)$ was of length 16 bits and only all $2^n$ strings for $n=8$ were produced, the strings in $D(5)$ have lengths from 1 to 49 bits (excluding lengths 42 and 46 that never occur) and include every possible string of length $l < 12$. Among the 12 bit strings, only two were not produced (000110100111 and 111001011000). Of $n=13, \ldots, 15$ about half the $2^n$ strings were produced (and therefore have frequency and complexity values). Fig. \ref{PropLengthD(5)} shows the proportion of $n$-long strings appearing in $D(5)$ outputs, for $n\in \{1, \ldots, 49\}$. 

\begin{figure}[h]
\centering
\includegraphics[width=8.7cm]{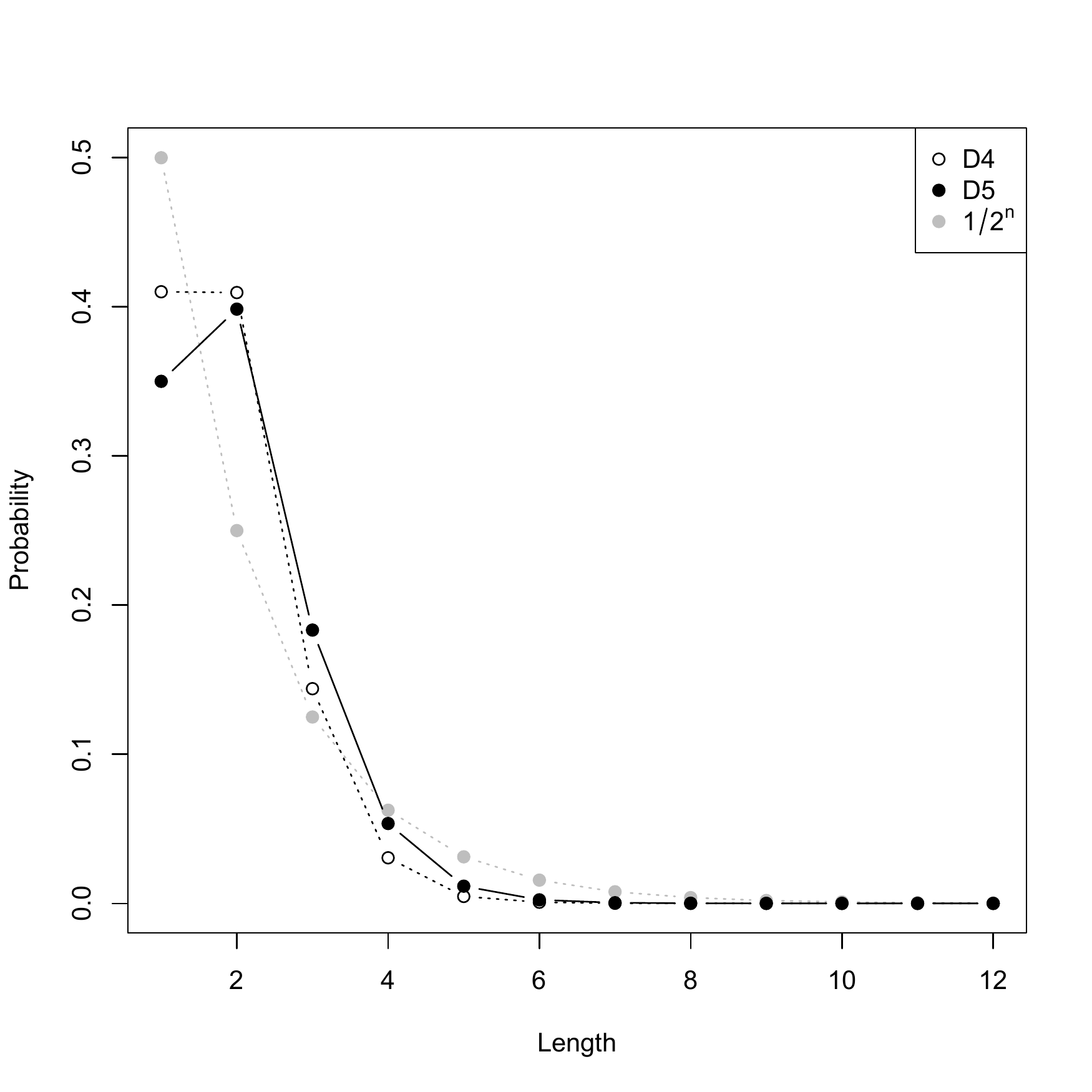}
\caption{\label{LevinD(4)D(5)} Cumulative probability of all $n$-long strings against $n$}
\end{figure}

The cumulative probability of every $n$-long string gives a probability law on $\mathbb{N}^*$. Fig. \ref{LevinD(4)D(5)} shows such a law obtained with $D(5)$, with $D(4)$, and with the theoretical $2^{-n}$ appearing in Levin's semi-measure. The most important difference may be the fact that this law does not decrease for $D(5)$, since length 2 is more likely than length 1.

\subsection*{Global simplicity}
Some binary sequences may seem simple from a global point of view because they show symmetry (1011 1101) or repetition (1011 1011). 
Let us consider the string $s=1011$ as an example. We have
$P_{D(5)}(s)=3.267414\times 10^{-3}$. The repetition $ss=10111011$ has
a much lower probability $P_{D(5)}(ss)=4.645999\times 10^{-7}$. This
is not surprising considering the fact that $ss$ is much longer than
$s$, but we may then wish to consider other strings based on $s$. In
what follows, we will consider three methods (repetition,
symmetrization, 0-complementation). The repetition of $s$ is
$ss=10111011$, the ``symmetrized" $s\bar{s}=1011 1101$, and the
0-complementation $10110000$. These three strings of identical length
have different probabilities ($4.645999\times10^{-7}$, $5.335785\times
10^{-7}$ and $3.649934\times 10^{-7}$ respectively). 

\begin{figure}[h]
\centering
\caption{\label{SymCompRep} Mean $\pm$ standard deviation of $D(5)$ of $2n$-long strings given by process of symmetrization (Sym), 0-complementation (Comp) and repetition (Rep) of all $n$-long strings. The dotted horizontal line shows the minimum mean among Sym, Comp and Rep. The density of $D(5)$ (smoothed with Gaussian kernel) for all $2n$-long strings is given in the right-margin.}
\includegraphics[width=12cm]{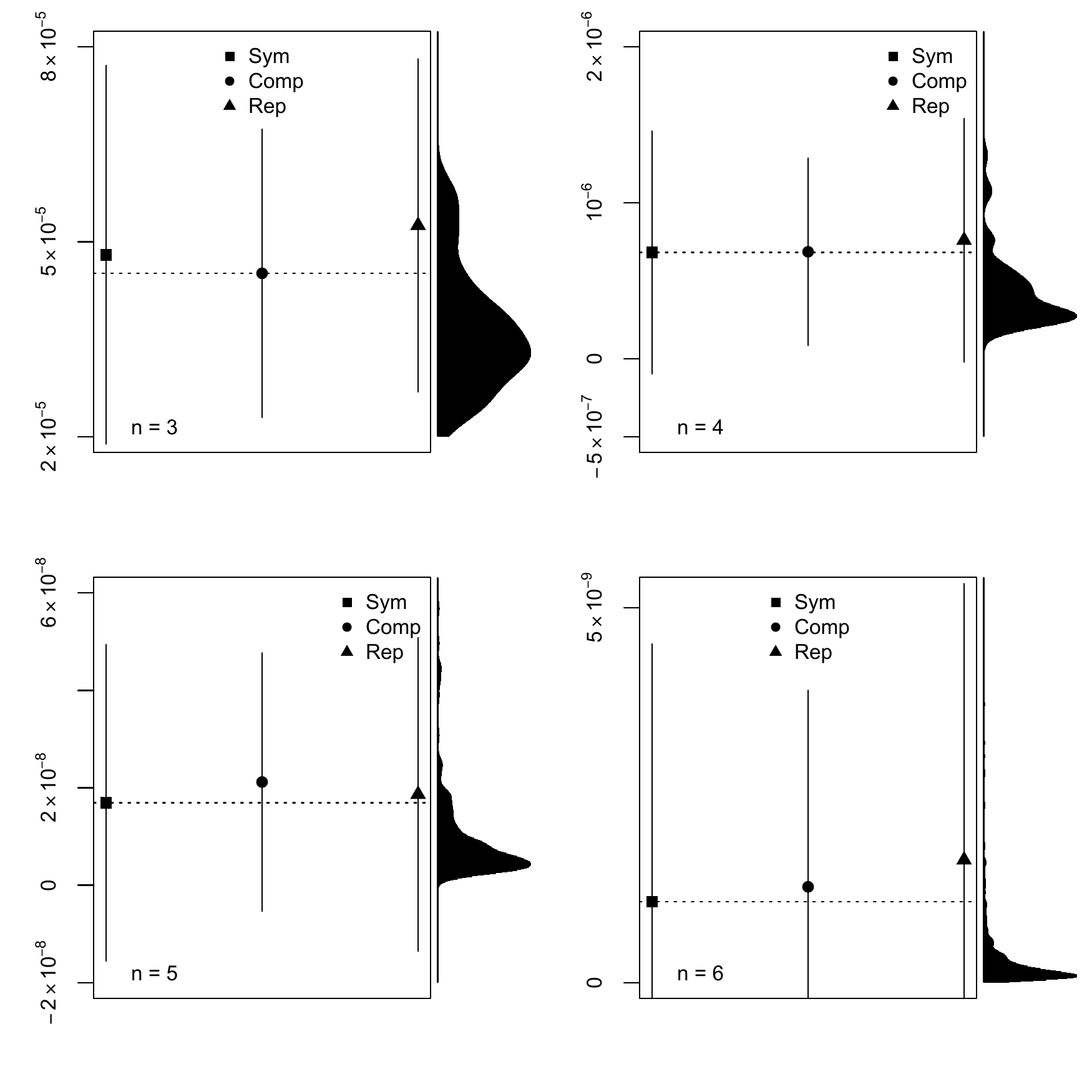}
\end{figure}

Let us now consider all strings of length 3 to 6, and their symmetrization, 0-complementation and repetition. Fig. \ref{SymCompRep} is a visual presentation of the results. In each case, even the minimum mean between the mean of symmetrized, complemented and repeated patterns (dotted horizontal line) lies in the upper tail of the $D(5)$ distribution for $2n$-length strings. And this is even more obvious with longer strings. Symmetry, complementation and repetition are, on average, recognized by $D(5)$.

Another method for finding ``simple" sequences is based on the fact that the length of a string is negatively linked to its probability of appearing in $D(5)$. When ordered by decreasing probability, strings show increasing lengths. Let's call those sequences for which length is greater than that of the next string ``climbers". The first 50 climbers appearing in $D(5)$ are given in Table \ref{climbers} and show subjectively simple patterns, as expected.

\begin{center}
\begin{table}[h]
   \caption{\label{climbers} Minimal examples of emergence: the first 50 climbers.}
   \begin{center}
      \tabcolsep=0.12cm
   \begin{tabular}{|c|c|c|c|c|}
   \hline
00000000 & 000000000 & 000000001 & 000010000 & 010101010\\
\hline
000000010 & 000000100 & 0000000000 & 0101010101 & 0000001010\\
\hline
0010101010 & 00000000000 & 0000000010 & 0000011010 & 0100010001\\
\hline
0000001000 & 0000101010 & 01010101010 & 0000000011 & 0101010110\\
\hline
0000000100 & 0000010101 & 000000000000 & 0000110000 & 0000110101\\
\hline
0000000110 & 0110110110 & 00000010000 & 0000001001 & 00000000001\\
\hline
0010101101 & 0101001001 & 0000011000 & 00010101010 & 01010010101\\
\hline
0010000001 & 00000100000 & 00101010101 & 00000000010 & 00000110000\\
\hline
00000000100 & 01000101010 & 01010101001 & 01001001001 & 010101010101\\
\hline
01001010010 & 000000000001 & 00000011000 & 00000000101 & 0000000000000\\
\hline
   \end{tabular}
   \end{center}
\end{table}
\end{center}

Strings are not sorted by length but follow an interesting distribution of length differences that agrees with our intuition of simplicity and randomness and is in keeping with the expectation from an approximation to $m(s)$ and therefore $K(s)$.

\subsection*{Binomial behavior}

\begin{figure}[h!]
\centering
\includegraphics[width=12cm]{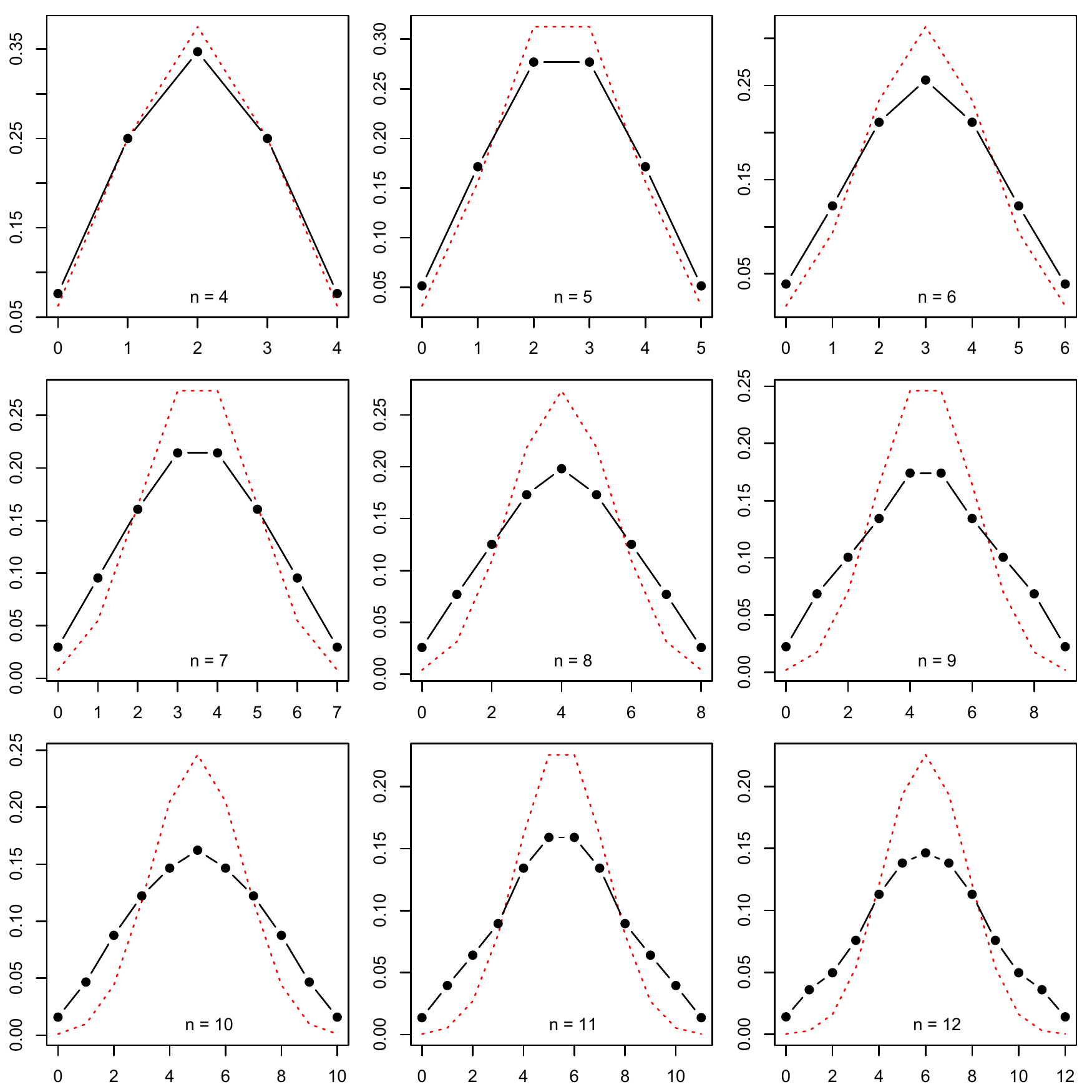}
\caption{\label{binomial} Distributions of the number of zeros in $n$-long binary sequences according to a truly random drawing (red, dotted), or a $D(5)$ drawing (black, solid) for length 4 to 12}
\end{figure}

In a random binary string of length $n$, the number of `0s' conforms to a binomial law given by $P(k)=2^{-n}{n \choose k}$. On the other hand, if a random Turing machine is drawn, simpler patterns are more likely to appear. Therefore, the distribution arising from Turing machine should be more scattered, since most simple patterns are often unbalanced (such as 0000000).  This is indeed what Fig.~\ref{binomial} shows: compared to truly random sequences of length $n$, $D(5)$ yields a larger standard deviation.

\subsection*{Bayesian application}

$D(5)$ allows us to determine, using a Bayesian approach, the probability that a given sequence is random: 
Let $s$ be a sequence of length $l$. This sequence may be produced by a machine, let's say a 5-state Turing machine (event $M$), or by a random process (event $R$). Let's set the prior probability at $P(R)=P(M)=\frac{1}{2}$.
 Because $s$ does not have a fixed length, we cannot use the usual probability $P(s)=\frac{1}{2^l}$, but we may, following Levin's idea, use $P(s|R)=\frac{1}{2^{2l}}$.
Given $s$, we can compute $$P(R|s)=\frac{P(s|R)P(R)}{P(s)},$$ with $$P(s)=P(s|M)P(M)+P(s|R)P(R)=\frac{P_{D(5)}(s)}{2}+\frac{1}{2^{2l+1}}.$$ Since $P(R)=\frac{1}{2}$ and $P(s|R)=\frac{1}{2^{2l}}$, the formula becomes $$P(R|s)=\frac{1}{{2^{2l}}P_{D(5)}(s)+1}.$$ 

There are 16 strings $s$ such that $P(R|s)<10^{-16}$ (the ``least random strings"). Their lengths lie in the range $[47,49]$. An example is: 11101110111011101110111011101110111011111010101. The fact that the ``least random" strings are long can intuitively be deemed correct: a sequence must be long before we can be certain it is not random. A simple sequence such as 00000000000000000000 (twenty `0s') gives a $P(R|s)=0.006$.

A total of 192 strings achieve a $P(R|s)>1-1.7\times 10^{-4}$. They all are of length 12 or 13. Examples are the strings 1110100001110, 1101110000110 or 1100101101000. This is consistent with our idea of a random sequence. However, the fact that only lengths 12 and 13 appear here may be due to the specificity of $D(5)$.

\subsection*{Comparing $D(4)$ and $D(5)$}

Every 4-state Turing machine may be modeled by a 5-state Turing machine whose fifth state is
never attained. Therefore, the 1832 strings produced by $D(4)$ calculated in~\cite{delahayezenil} also appear in $D(5)$. We thus have 1832 ranked elements in $D(4)$ to compare with. The basic idea at the root of this work is that $D(5)$ is a refinement (and major extension) of $D(4)$, previously calculated in an attempt to understand and evaluate algorithmic complexity. This would be hopeless if $D(4)$ and $D(5)$ led to totally different measures and rankings of simplicity versus complexity (randomness).

\subsection*{Agreement in probability}

\begin{figure}[h!]
\centering
\includegraphics[width=14.5cm]{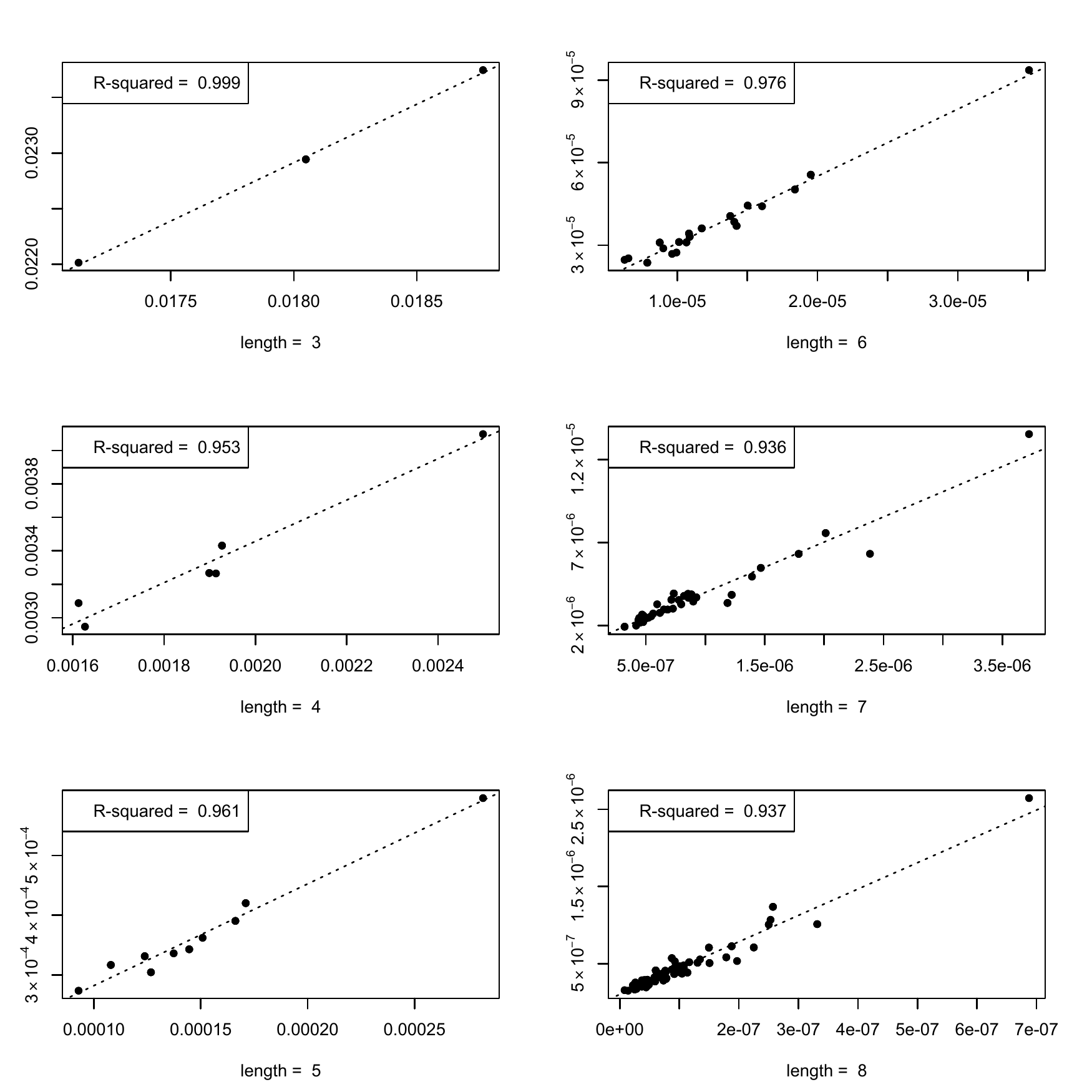}
\caption{\label{corD(4)D(5)} ${D(5)}$ against ${D(4)}$, for $n$-long strings}
\end{figure}

The link between ${D(4)}$ and ${D(5)}$ seen as measures of simplicity may be measured by the determination coefficient $r^2$, $r$ being the Pearson correlation coefficient. This coefficient is $r^2=99.23\%$, which may be understood as ``${D(4)}$ explains 99.23\% of the variations of ${D(5)}$". 
The scatterplot in Fig. \ref{corD(4)D(5)} displays ${D(5)}(s)$ against ${D(4)}(s)$ for all strings $s$ of length $n=3, \ldots, 8$ (8 being the largest integer $l$ such that $D(4)$ comprises every $l$-long sequence). 

The agreement between ${D(5)}$ and ${D(4)}$ is almost perfect, but there are still some differences. Possible outliers may be found using a studentized residual in the linear regression of ${D(5)}$ against ${D(4)}$. 
The only strings giving absolute studentized residuals above 20 are 0 and 1. The only strings giving absolute studentized residuals lying between 5 and 20 are all the $3$-long strings. All $4$-long strings fall between 2 and 5.
 This shows that the differences between ${D(5)}$ and ${D(4)}$ may be explained by the relative importance given to the diverse lengths, as shown above (Fig.~\ref{LevinD(4)D(5)}).

\subsection*{Agreement in rank}
\label{arank}

\begin{figure}[h]
\centering
\includegraphics[width=9cm]{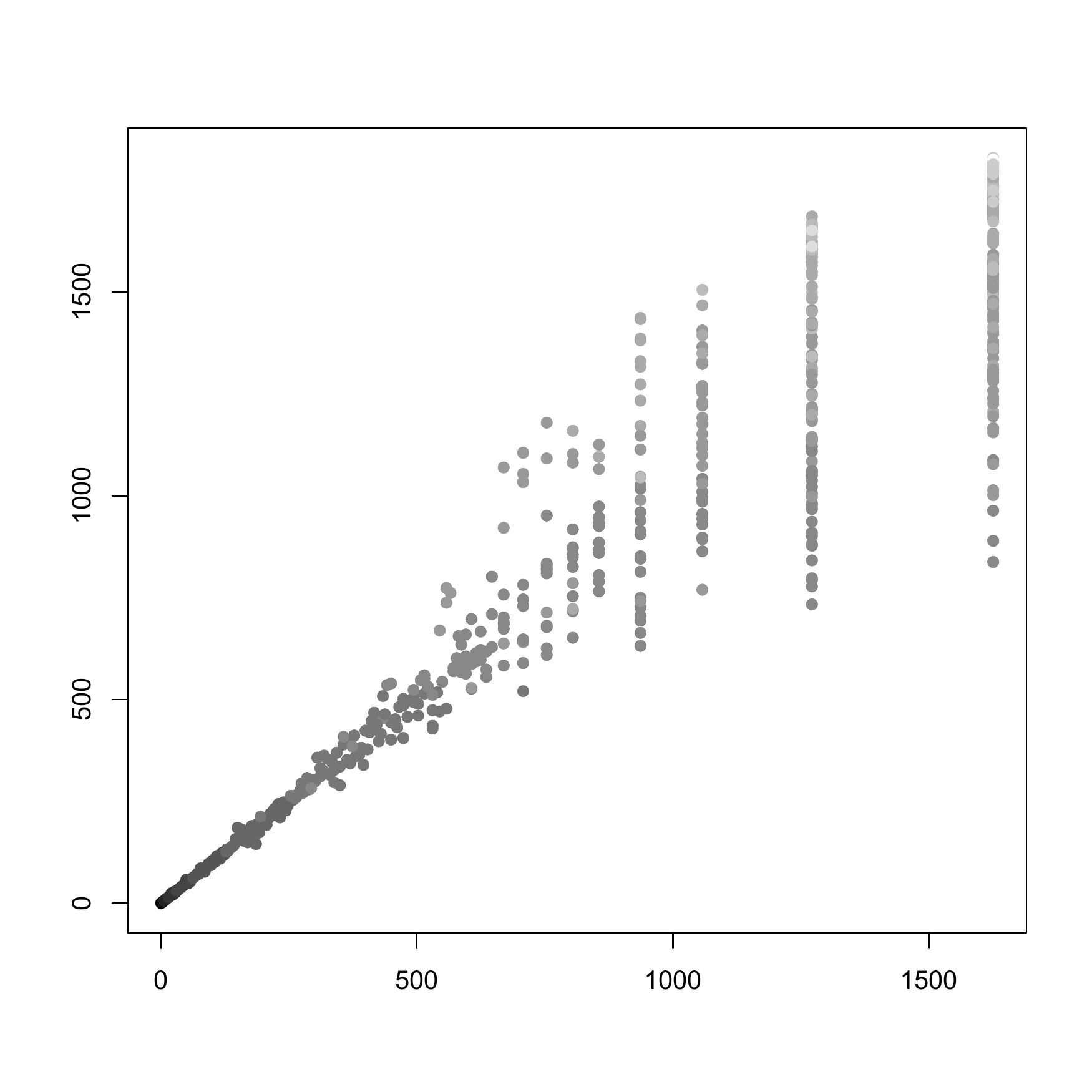}
\caption{\label{rankD(4)D(5)} $R_5$ (rank according to ${D(5)}$) against $R_4$. The grayscale indicates the length of the strings: the darker the point, the shorter the string.}
\end{figure}

There are some discrepancies between ${D(5)}$ and ${D(4)}$ due to length effects. Another way of studying the relationship between the two measures is to turn our attention to ranks arising from ${D(5)}$ and ${D(4)}$. The Spearman coefficient is an efficient tool for comparing ranks. 
Each string may be associated with a rank according to decreasing values of ${D(5)}$ ($R_5$) or ${D(4)}$ ($R_4$). A greater rank means that the string is less probable. Fig. \ref{rankD(4)D(5)} displays a scatterplot of ranks according to ${D(5)}$ as a function of ${D(4)}$-rank. Visual inspection shows that the ranks are similar, especially for shorter sequences. The Spearman correlation coefficient amounts to 0.9305, indicating strong agreement. 

Not all strings are equally ranked and it may be interesting to take a closer look at outliers. Table~\ref{largest} shows the 20 strings for which $|R_4-R_5|\geq 600$. All these sequences are ties in $D(4)$, whereas $D(5)$ distinguishes 5 groups. Each group is made up of 4 equivalent strings formed from simple transformations (reversing and complementation). This confirms that $D(5)$ is fine-grained compared to $D(4)$.

The shortest sequences such that $|R_4-R_5|\geq 5$ are of length 6. Some of them show an intriguing pattern, with an inversion in ranks, such as 000100 ($R_5=85, R_4=77$) and 101001 with reversed ranks.

\begin{center}
\begin{table}[h]
   \caption{\label{largest} The 20 strings for which $|R_4-R_5|\geq 600$}
   \begin{center}
      \tabcolsep=0.12cm
   \begin{tabular}{|c|c|c|}
\hline
   sequence & $R_4$ & $R_5$\\
\hline
010111110 & 1625.5 & 837.5\\011111010 & 1625.5 & 837.5\\100000101 & 1625.5 & 837.5\\101000001 & 1625.5 & 837.5\\000011001 & 1625.5 & 889.5\\011001111 & 1625.5 & 889.5\\100110000 & 1625.5 & 889.5\\111100110 & 1625.5 & 889.5\\001111101 & 1625.5 & 963.5\\010000011 & 1625.5 & 963.5\\101111100 & 1625.5 & 963.5\\110000010 & 1625.5 & 963.5\\0101010110 & 1625.5 & 1001.5\\0110101010 & 1625.5 & 1001.5\\1001010101 & 1625.5 & 1001.5\\1010101001 & 1625.5 & 1001.5\\0000000100 & 1625.5 & 1013.5\\0010000000 & 1625.5 & 1013.5\\1101111111 & 1625.5 & 1013.5\\1111111011 & 1625.5 & 1013.5\\
\hline
   \end{tabular}
   \end{center}
\end{table}
\end{center}

On the whole, ${D(5)}$ and ${D(4)}$ are similar measures of simplicity, both from a measurement point of view and a ranking point of view. Some differences may arise from the fact that $D(5)$ is more fine-grained than $D(4)$. Other unexpected discrepancies still remain: we must be aware that $D(5)$ and $D(4)$ are both approximations of a more general limit measure of simplicity versus randomness. Differences are inevitable, but the discrepancies are rare enough to allow us to hope that $D(5)$ is for the most part a good approximation of this properties.

\subsection*{Kolmogorov complexity approximation}

It is now straightforward to apply the Coding theorem to convert string frequency (as a numerical approximation of algorithmic probability $\mathfrak{m}(s)$) to estimate an evaluation of Kolmogorov complexity (see Tables~\ref{trans} and~\ref{trans2}) for the Turing machine formalism chosen. Based on the concept of algorithmic probability we define:

\begin{center}
$K_{D(n)}(s)=-\log_2 {D(n)(s)}$
\end{center}

\begin{center}
\begin{table}[h]
   \caption{\label{trans} Top 20 strings in $D(5)$ with highest frequency and therefore lowest Kolmogorov (program-size) complexity. From frequency (middle column) to complexity (extreme right column) applying the coding theorem in order to get $K_{D(5)}$ which we will call $K_{\mathfrak{m}(s)}$ as our current best approximation to an experimental $\mathfrak{m}(s)$, that is $D(5)$, through the Coding theorem.}
   \begin{center}
      \tabcolsep=0.12cm
   \begin{tabular}{|c|c|c|}
\hline
sequence & frequency ($\mathfrak{m}(s)$) & complexity ($K_\mathfrak{m}(s)$) \\
\hline
 1&0.175036&2.51428\\
 0&0.175036&2.51428\\
 11&0.0996187&3.32744\\
 10&0.0996187&3.32744\\
 01&0.0996187&3.32744\\
 00&0.0996187&3.32744\\
 111&0.0237456&5.3962\\
 000&0.0237456&5.3962\\
 110&0.0229434&5.44578\\
 100&0.0229434&5.44578\\
 011&0.0229434&5.44578\\
 001&0.0229434&5.44578\\
 101&0.0220148&5.50538\\
 010&0.0220148&5.50538\\
 1111&0.0040981&7.93083\\
 0000&0.0040981&7.93083\\
 1110&0.00343136&8.187\\
 1000&0.00343136&8.187\\
 0111&0.00343136&8.187\\
 0001&0.00343136&8.187\\
\hline
   \end{tabular}
   \end{center}
\end{table}
\end{center}

First it is worth noting that the calculated complexity values in Tables~\ref{trans} and~\ref{trans2} are real numbers, when they are supposed to be the lengths of programs (in bits) that produce the strings, hence integers. The obvious thing to do is to round the values to the next closest integer, but this would have a negative impact as the decimal expansion provides a finer classification. Hence the finer structure of the classification is favored over the exact interpretation of the values as lengths of computer programs. It is also worth mentioning that the lengths of the strings (as shown in Table~\ref{trans2}) are almost always smaller than their Kolmogorov (program-size) values, which is somehow to be expected from this approach. Consider the single bit. It not only encodes itself, but the length of the string (1 bit) as well, because it is produced by a Turing machine that has reached the halting state and produced this output upon halting. 


Also worth noting is the fact that the strings 00, 01, 10 and 11 all have the same complexity, according to our calculations (this is the case from $D(2)$ to $D(5)$). It might just be the case that the strings are too short to really have different complexities, and that a Turing machine that can produce one or the other is of exactly the same length. To us the string 00 may look more simple than 01, but we do not have many arguments to validate this intuition for such short strings, and it may be an indication that such intuition is misguided (think in natural language, if spelled out in words, 00 does not seem to have a much shorter description than the shortest description of 01).

Compare this phenomenon of program-sizes being greater than the length of these short strings to the extent of the problem posed by compression algorithms, which collapse all different strings of up to some length (usually around 100 bits), retrieving the same complexity approximation for all of them despite their differences. One way to overcome this minor inconvenience involved in using the alternative approach developed here is to subtract a constant (no greater than the smallest complexity value) from all the complexity values, which gives these strings lower absolute random complexity values (preserving the relative order). But even if left ``random", this alternative technique can be used to distinguish and compare them, unlike the lossless compression approach that is unable to further compress short strings.

\begin{center}
\begin{table}[h]
   \caption{\label{trans2} 20 random strings (sorted from lowest to highest complexity values) from the first half of $D(5)$ to which the coding theorem has been applied (extreme right column) to approximate $K(s)$.}
   \begin{center}
      \tabcolsep=0.12cm
   \begin{tabular}{|c|c|c|}
\hline
   string length & string & complexity ($K_\mathfrak{m}(s)$) \\
\hline
11&11011011010&28.1839 \\
12&101101110011&32.1101 \\
12&110101001000&32.1816 \\
13&0101010000010&32.8155 \\
14&11111111100010&34.1572 \\
12&011100100011&34.6045 \\
15&001000010101010&35.2569 \\
16&0101100000000000&35.6047 \\
13&0110011101101&35.8943 \\
15&101011000100010&35.8943 \\
16&1111101010111111&25.1313 \\
18&000000000101000000&36.2568 \\
15&001010010000000&36.7423 \\
15&101011000001100&36.7423 \\
17&10010011010010011&37.0641 \\
21&100110000000110111011&37.0641 \\
14&11000010000101&37.0641 \\
17&01010000101101101&37.4792 \\
29&01011101111100011101111010101&37.4792 \\
14&11111110011110&37.4792 \\
\hline
   \end{tabular}
   \end{center}
\end{table}
\end{center}

The phenomenon of complexity values greater than the lengths of the strings is transitional. Out of the 99,608 strings in $D(5)$, 212 have greater string lengths than program-size values. The first string to have a smaller program-size value than string length is the string 10101010101010101010
101010101010101010101 (and its complementation), of length 41 but program-size of 33.11 (34 if rounded). The mean of the strings with greater program-size than length is 38.35,  The string with the greatest difference between length and program-size in $D(5)$ are strings of low Kolmogorov complexity such as 0101010001000100010001000100010001000100010001010, of length 49 but with an approximated Kolmogorov complexity (program-size) value of 39.06. Hence far from random, both in terms of the measure and in terms of the string's appearance. 

\subsection*{Randomness in $D(5)$}

Paradoxically, the strings at the bottom of $D(5)$ as sorted from highest to lowest frequency and therefore lowest to highest Kolmogorov (random) complexity are not very random looking, but this is to be expected, as the actual most random strings of these lengths would have had very low frequencies and would not therefore have been produced. In fact what we are looking at the bottom are some of the longest strings with greatest structure (and hence with lowest Kolmogorov complexity) in $D(5)$. Table~\ref{transb2}, however, shows the bottom of the length $n=12$ classification extracted from $D(5)$, for which all, but 2, $2^n$ binary strings were produced, hence displaying more apparent randomness in the subset of length $n=12$ but still likely unstable due to the fact that they were produced by a relatively small number of Turing machines (in fact 2 strings of length 12 were never produced), compared to the ones at the top for $n=12$ that, as we have shown, can be expected to be stable (from comparing $D(4)$ and $D(5)$).

\begin{center}
\begin{table}[h]
   \caption{\label{transb2} Bottom 21 strings of length $n=12$ with smallest frequency in $D(5)$.}
   \begin{center}
      \tabcolsep=0.12cm
   \begin{tabular}{|c|c|c|}
\hline
100111000110&100101110001&100011101001\\
\hline
100011100001&100001110001&011110001110\\
\hline
011100011110&011100010110&011010001110\\
\hline
011000111001&000100110111&111000110100\\
\hline
110100111000&001011000111&000111001011\\
\hline
110100011100&110001110100&001110001011\\
\hline
001011100011&110000111100&001111000011\\
\hline
   \end{tabular}
   \end{center}
\end{table}
\end{center}

\subsection*{Robustness of $K_{D(n)}$}

An important question is how robust is $K_{D(n)}$, that is how sensitive it is to $n$. We know that the invariance theorem guarantees that the values converge in the long term, but the invariance theorem tells nothing about the rate of convergence. We have shown that $D(n+1)$ respects the order of $D(n)$ except for very few and minor value discrepancies concerning the least frequent strings (and therefore the most unstable given the few machines generating them). This is not obvious despite the fact that all Turing machines with $n$ states in $(n, m)$ are included in the space of $(n+1, m)$ machines (that is, the machines that never reach one of the $n+1$ states), because the number of machines in $(n+1, m)$ overcomes by far the number of machines in $(n, m)$, and a completely different result could have been then produced. However, the agreement between $D(n)$ and $D(n+1)$ seems to be similarly high among, and despite, the few cases $n\leq6$ in hand to compare with. The only way for this behaviour to radically change for $n>5$ is if for some $n^\prime$, $D(n^\prime)$ starts diverging in ranks from $D(n^\prime-1)$ on before starting to converge again (by the invariance theorem). If one does not have any reason to believe in such a change of behavior, the rate of rank convergence of $D(n)$ is close to optimal very soon, even for the relatively ``small" sets of Turing machines for small $n$.

One may ask how robust the complexity values and classifications may be in the face of changes in computational formalism (e.g. Turing machines with several tapes, and all possible variations). We have shown~\cite{zenilalgo} that radical changes to the computing model produce reasonable (and correlated with different degrees of confidence) ranking distributions of complexity values (using even completely different computing models, such as unidimensional deterministic cellular automata and Post tag systems).

We have also calculated the maximum differences between the Kolmogorov complexity evaluations of the strings occurring in every 2 distributions $D(n)$ and $D(n+1)$ for $n= 2, \ldots, 4$. This provides estimations for the constant $c$ in the invariance theorem (Eq.~\ref{invariance}) determining the maximum difference in bits among all the strings evaluated with one or another distribution, hence shedding light on the robustness of the evaluations under this procedure. The smaller the values of $c$ the more stable our method. The values of these bounding constants (in bits) among the different numerical evaluations of $K$ using $D(n)$ for $n=2, \dots, 5$ after application of the Coding theorem (Eq.~\ref{codingeq}) are:

\begin{center}
$| K_{D(2)}(s) - K_{D(3)}(s) | \leq c = 4.090; 4.090; 3.448; 0.39$ \\
$| K_{D(3)}(s) - K_{D(4)}(s) | \leq c = 4.10; 3.234; 2.327; 2.327$\\
$| K_{D(4)}(s) - K_{D(5)}(s) | \leq c = 5.022; 4.274; 3.40; 2.797$\\
\end{center}

Where $K_{D(n)}(s)$ means $K(s)$ evaluated using the output frequency distribution $D(n)$ after application of the Coding theorem (Eq.~\ref{codingeq}) for $n= 2, \ldots, 5$ ($n=1$ is a trivial non interesting case) and where every value of $c$ is calculated by quartiles (separated by semicolons), that is, the calculation of $c$ among all the strings in the 2 compared distributions, then among the top 3/4, then the top half and finally the top quarter by rank. Notice that the estimation of $c$ between $D(2)$ and $D(3)$, and $D(3)$ and $D(4)$ remained almost the same among all strings occurring in both, at about 4 bits. This means one could write a ``compiler" (or translator) among the two distributions for all their occurring strings of size only 4 bits providing one or the other complexity value for $K$ based on one or the other distribution. The differences are considerably smaller for more stable strings (towards the top of the distributions). One may think that given that the strings with their occurrences in $D(n+1)$ necessarily contain those in $D(n)$ for all $n$ (because the space of all Turing machines with an additional state always contain the computations of the Turing machines will less states), the agreement should be expected. However, the contribution of $D(n)$ to $D(n+1)$ contributes with $\log$ the number of strings in $D(n+1)$. For example, $D(4)$ contributes only 1832 strings to the 99\,608 produced in $D(5)$ (that is less than 2\%). All in all, the largest difference found between $D(4)$ and $D(5)$ is only of 5 bits of among all the strings occurring both in $D(4)$ and $D(5)$ (1832 strings), where the values of $K$ in $D(4)$ are between 2.285 and 29.9.

\section*{Concluding remarks}

We have put forward a method based on algorithmic probability that produces frequency distributions based on the production of strings using a standard (Rado's) model of Turing machines generally used for the Busy Beaver problem. The distributions show very small variations, being the result of an operation that makes incremental changes based on a very large number of calculations having as consequence the production of stable numerical approximations of Kolmogorov complexity for short strings for which error estimations of $c$ from the invariance theorem were also estimated. Any substantial improvement on $D(5)$, for example, by approximation of a $D(n)$ for $n>5$, is unlikely to happen with the current technology as the number of Turing machines grows exponentially in the number of states $n$. However, we have shown here based both on theoretical and experimental grounds that one can choose informed runtimes significantly smaller than that of the Busy Beaver bound and capture most of the output determining the output frequency distribution. An increase of computational power by, say, one order of magnitude will only deliver a linear improvement on $D(5)$.

The experimental method presented is computationally expensive, but it does not need to be executed more but once for a set of (short) strings. As a result this can now be considered an alternative to lossless compression as a complementary technique for approximating Kolmogorov complexity. An \emph{Online Algorithmic Complexity Calculator} (OACC) implementing this technique and releasing the data for public use has been made available at \url{http://www.complexitycalculator.com}. 

The data produced for this paper has already been used in connection to graph theory and complex networks~\cite{zenilgraphs}, showing, for example, that it produces better approximations of Kolmogorov complexity of small graphs (by comparing it to their duals) than lossless compressibility. In~\cite{2D} it is also shown how the method can be used to classify images and space-time diagrams of dynamical systems, where its results are also compared to the approximations obtained using compression algorithms, with which they show spectacular agreement. In~\cite{chaos}, it is used to investigate the ratios of complexity in rule spaces of cellular automata of increasing size, supported by results from block entropy and lossless compressibility. In~\cite{gauvrit}, it is also used as a tool to assess subjective randomness in the context of psychometrics. Finally in~\cite{computability}, the method is used in numerical approximations to another seminal measure of complexity (Bennett's Logical Depth), where it is also shown to be compatible with a calculation of strict (integer-value) program-size complexity as measured by an alternative means (i.e. other than compression). The procedure promises to be a sound alternative, bringing theory and practice into alignment and constituting evidence that confirms the possible real-world applicability of Levin's distribution and Solomonoff's universal induction (hence validating the theory itself, which has been subject to criticism largely on grounds of simplicity bias and inapplicability). As Gregory Chaitin has pointed out~\cite{thesisreport} when commenting on this very method of ours:

\begin{quote}
The theory of algorithmic complexity is of course now widely accepted, but was initially rejected by many because of the fact that algorithmic complexity depends on the choice of universal Turing machine and short binary sequences cannot be usefully discussed from the perspective of algorithmic complexity. \ldots discovered employing [t]his empirical, experimental approach, the fact that most reasonable choices of formalisms for describing short sequences of bits give consistent measures of algorithmic complexity! So the dreaded theoretical hole in the foundations of algorithmic complexity turns out, in practice, not to be as serious as was previously assumed. \ldots [hence, of this approach] constituting a marked turn in the field of algorithmic complexity from deep theory to practical applications.
\end{quote}

This also refers to the fact that we have found an important agreement in distribution---and therefore of estimations of Kolmogorov complexity upon application of the algorithmic Coding theorem---with other abstract computing formalisms such as one-dimensional cellular automata and Post's tag systems~\cite{zenilalgo}. In this paper we have provided strong evidence that the estimation and scaling (albeit limited by computational power) of the method is robust and much less dependent on formalism and size sample than what originally could have been anticipated by the invariance theorem.








\end{document}